\documentclass[conference]{IEEEtran}
\IEEEoverridecommandlockouts
\usepackage{dblfloatfix}
\usepackage{setspace}
\usepackage{adjustbox}
\usepackage{cite}
\usepackage{amsmath,amssymb,amsfonts, pifont}
\usepackage{algorithmic}
\usepackage{graphicx}
\usepackage{textcomp}
\usepackage{xcolor}
\usepackage{listings}
\usepackage{color}
\usepackage{anyfontsize}

\definecolor{lightgray}{rgb}{9, 9, .9}
\definecolor{darkgray}{rgb}{.4, .4, .4}
\definecolor{purple}{rgb}{0.65, 0.12, 0.82}
\definecolor{gray}{rgb}{0.4,0.4,0.4}
\definecolor{line-numbers}{rgb}{0.4,0.4,0.4}
\definecolor{tags}{rgb}{1, 0, 0}
\definecolor{darkblue}{rgb}{0.0,0.0,0.6}
\definecolor{cyan}{rgb}{0.0,0.6,0.6}
\lstdefinelanguage{Android}
{
  sensitive=true,%
    morecomment=[l]//,%
  morecomment=[s]{/}{/},%
  morestring=[b]",%
  morestring=[b]',%
  stringstyle=\color{black},
  identifierstyle=\color{darkblue},
  keywordstyle=\color{cyan},
  morekeywords={abstract,any,as,boolean,break,byte,case,catch,char,
  class, const,continue,def,default,do,double,else,extends,false,final,finally, float,for,goto,if,implements,import,instanceof,in,int,interface,label, long,native,new,null,package,private,protected,public,return,short, static,strictfp,super,switch,synchronized,this,throw,throws,transient, true,try,void,volatile,while,with}
}
\usepackage[T1]{fontenc}
\usepackage{listings,xcolor,beramono}
\usepackage{tikz}
\makeatletter
\newenvironment{btHighlight}[1][]
{\begingroup\tikzset{bt@Highlight@par/.style={#1}}\begin{lrbox}{\@tempboxa}}
{\end{lrbox}\bt@HL@box[bt@Highlight@par]{\@tempboxa}\endgroup}

\newcommand\btHL[1][]{%
  \begin{btHighlight}[#1]\bgroup\aftergroup\bt@HL@endenv%
}
\def\bt@HL@endenv{%
  \end{btHighlight}%
  \egroup
}
\newcommand{\bt@HL@box}[2][]{%
  \tikz[#1]{%
    \pgfpathrectangle{\pgfpoint{1pt}{0pt}}{\pgfpoint{\wd #2}{\ht #2}}%
    \pgfusepath{use as bounding box}%
    \node[anchor=base west, fill=orange!30,outer sep=0pt,inner xsep=1pt, inner ysep=0pt, rounded corners=3pt, minimum height=\ht\strutbox+1pt,#1]{\raisebox{1pt}{\strut}\strut\usebox{#2}};
  }%
}
\makeatother

\lstdefinestyle{Android}{
    language={Android}, 
     moredelim=**[is][\btHL]{`}{`},
    moredelim=**[is][{\btHL[fill=green!30]}]{*}{*},
    moredelim=**[is][{\btHL[fill=cyan!30]}]{~}{~},
    extendedchars=true,
        }
\usepackage{qtree}
\usepackage{multicol}        
\def\BibTeX{{\rm B\kern-.05em{\sc i\kern-.025em b}\kern-.08em
    T\kern-.1667em\lower.7ex\hbox{E}\kern-.125emX}}

\usepackage{balance}
\begin{document}
\author{\IEEEauthorblockN{Abdul Moiz, Manar H. Alalfi}
\IEEEauthorblockA{\textit{Department of Computer Science} \\
\textit{Ryerson University,
Toronto, ON, Canada} \\
{\{a1moiz,manar.alalfi\}@ryerson.ca}}}
\title{An Approach for the Identification of Information Leakage in Automotive Infotainment systems
}
\vspace{-0.7 cm}
%
\maketitle
\vspace{-0.3 cm}
\begin{abstract}
The advancements in the digitization world has revolutionized the automotive industry. Today's modern cars are equipped with internet, computers that can provide autonomous driving functionalities as well as infotainment systems that can run mobile operating systems, like Android Auto and Apple CarPlay. Android Automotive is Google's android operating system tailored to run natively on vehicle's infotainment systems, it allows third party apps to be installed and run on vehicle's infotainment systems. Such apps may raise security concerns related to user's safety, security and privacy. This paper investigates security concerns of in-vehicle apps, specifically, those related to inter component communication (ICC) among these apps. ICC allows apps to share information via inter or intra apps components through a messaging object called intent. In case of insecure communication, Intent can be hijacked or spoofed by malicious apps and user’s sensitive information can be leaked to hacker’s database. We investigate the attack surface and vulnerabilities in these apps and provide a static analysis approach and a tool to find data leakage vulnerabilities. The approach can also provide hints to mitigate these leaks. We evaluate our approach by analyzing a set of Android Auto apps downloaded from Google Play store, and we report our validated results on vulnerabilities identified on those apps.
\end{abstract}
\section{Introduction}
\vspace{-0.15cm}
In recent years, the automotive industry has grown exponentially and the In-vehicle Infotainment (IVI) systems are equipped with Android Auto or Apple CarPlay, which provide functionalities to mirror mobile screen to IVI dashboard. Hence, enabling users to use their favourite mobile apps in their vehicles. Currently, these apps are limited to mobile devices capability and apps like, navigation, music, radio, making phone calls and read/send SMS. Google’s Android Automotive is an Android operating system tailored to run natively on vehicles IVI systems without any mobile device connection. Since it’s announcement in March 2017 by Google, car manufacturers; Polestar (Volvo's electric performance car), Renault–Nissan–Mitsubishi Alliance, General Motors, and Groupe PSA have announced to use Android Automotive to power the IVI systems in their Cars starting from 2020/2021. It will also enable the third party apps to install on IVI systems. Therefore, it will be crucial to analyze these apps for user’s safety, security and privacy. This justifies our focus to investigate Android Automotive apps.

Android Automotive OS contains car specific Application Program Interface (APIs) that Original Equipment Manufacturers (OEM) can use to enhance their application’s features. Some of these APIs can also be utilized by third party apps. Car information like, make, model, year, driving state, vehicle speed, location, and many more properties can be accessed by these apps. Therefore, it is important to understand how this information can be utilized by malicious apps.
\section{Background}
\vspace{-0.15cm}
In Android, each app exists in it’s own sandbox environment which can not be accessed by other apps. However, to communicate with app’s components, within the same or between different app, programmers rely on Inter Component Communication (ICC). ICC provides a communication mechanism that can send or broadcast a message component called \textit{intent} which can be received by activity, service or broadcast receiver. In case of a broadcast, \textit{intent} is sent to all app’s components installed on the device.
\subsection{Information Leakage Vulnerability}
Static analysis inspects program code to derive information about program’s behavior without running it. It uses abstraction and makes conservative assumptions about all possible program execution statically. Static analysis can be used to check for programming errors and security flaws. It is also used in modern compilers for optimization purposes. Static analysis approaches can be classified as type systems, or data-flow based approaches\cite{10.5555/1965094}.

A special type of data flow analysis is called Taint analysis, since it tracks the flow of sensitive “tainted” objects from sources to sinks. Where sources represents variables that receive user's controllable inputs, such as password, and sinks represents functions that communicate information it receive to an external component such as the internet. As such, taint analysis can be used to identify information leakage and other types of program vulnerabilities.

In Android apps, there are many ways for a source to end up in a sink. In this paper, our focus is on \textit{Intent} as a source and \textit{sendBroadcast} as a sink. Listing \ref{lst1} presents an example where an \textit{Intent} is created on line 176. Sensitive car information is added to the \textit{Intent} in (line 179-184). This information is then broadcasted to all apps (line 186). This example, demonstrates the case where any malicious app can eavesdrop on this \textit{Intent} and send it's sensitive information to a malicious server. However, the problem can be mitigated if security best practices are implemented. Listing \ref{lst1} (line 188 or 190) presents two possible solutions to mitigate the above leak. \textit{LocalBroadcastManager} is responsible to broadcast \textit{Intent} within the same app's components, i.e: activities, broadcast receivers and services. And \textit{sendBroadcast} with custom permission on (line 190) ensures that the \textit{Intent} will only be received by intra applications that have the same custom permissions defined in their app’s manifest file. Other applications which don’t have that same custom permission defined will not be able to receive this intent.
\vspace{-0.4cm}
\begin{lstlisting}[style=Android, firstnumber=175, caption={Sources and Sink}, label={lst1}]
private void broadcastIntent() {
`Intent intent = createIntent(key, value);`
intent.setAction("com.example.intent.broadcast");
CarInfoManager carInfo = (CarInfoManager) car.getCarManager(Car.INFO_SERVICE);
intent.putExtra("Battery Capacity", carInfo.getEvBatteryCapacity());
intent.putExtra("Connector Types", carInfo.getEvConnectorTypes());
intent.putExtra("Fuel Capacity", carInfo.getFuelCapacity());
intent.putExtra("Fuel Types", carInfo.getFuelTypes());
intent.putExtra("Manufacturer", carInfo.getManufacturer());
intent.putExtra("Model", carInfo.getModel());
    // SINK
~sendBroadcast(intent)~;
    // Solution 1
*LocalBroadcastManager.getInstance(this).sendBroadcast(v1);*
    // Solution 2
*sendBroadcast(intent,"com.intent.broadcast.permission");*}
\end{lstlisting}
\subsection{Threat Model}
 In this section, we present the threat models that represents the insecure usage of ICCs which can be utilised by malicious apps, i.e: \textit{intent hijacking, intent spoofing, intent eavesdropping} and \textit{intent collusion} attacks.
\subsubsection{Intent Hijacking}
As depicted in Figure \ref{fig:Tmodel}; App1’s component-B is expecting an \textit{intent} from component-A. However, by setting the attributes of intent filter same as component-B; malware's component-D is presenting itself as an authorized receiver of the \textit{intent} and as a result it can potentially leak the information. If the \textit{intent} contains any sensitive information that requires special permission (e.g: location, contact, vehicle speed) and if it doesn't restrict the receivers for permissions, then malware1 will receive the sensitive information without the need of permission. In this way, malware1 is escalating it’s privileges by \textit{hijacking} the \textit{intent}.
\subsubsection{Intent Spoofing}
As depicted in Figure \ref{fig:Tmodel}; If a malware app knows that component-B in App1 expects to receive an \textit{intent} from component-A and doesn't checks for permissions or attributes then malware1 can pretend to be component-A and can send the \textit{intent} to component-B which could trigger the corresponding actions on component-B. As a result this can also cause damage by crashing the app or stealing the information by requesting it from the sent \textit{intent}. 
\subsubsection{Intent Eavesdropping} 
\begin{figure}[t!]
\centerline{\includegraphics[width=.4\textwidth]{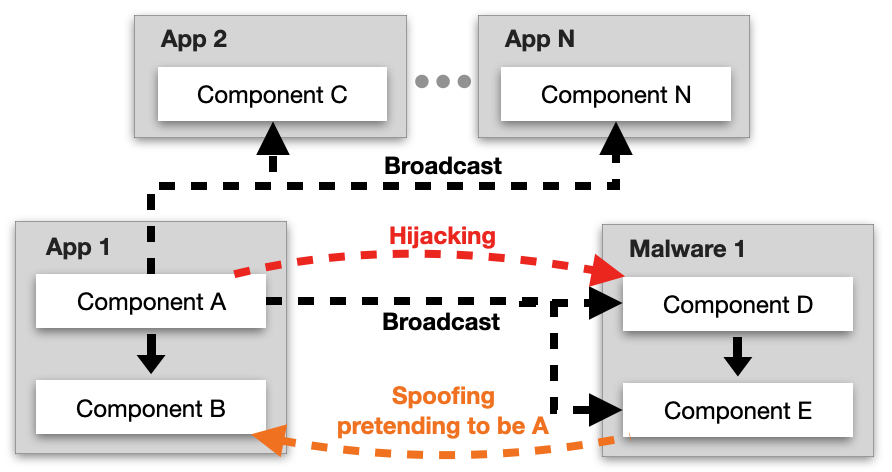}}
\caption{The intent hijacking, spoofing, eavesdropping attack model}
\vspace{-0.3 cm}
\label{fig:Tmodel}
\end{figure}

As depicted in Figure \ref{fig:Tmodel}, App1’s component-A is broadcasting an \textit{intent} which is received by all the app’s components. In this scenario, malicious apps will receive this \textit{intent} and as a result, it can send this information to malicious databases. If \textit{intent} contains any sensitive information that requires special permission (e.g: location, contact, vehicle speed) and if it doesn't restrict the receivers for permissions then malware1 will receive the sensitive information without the need of permission. In this way, malware1 is escalating it’s privileges by eavesdropping on the incoming \textit{intent}. 
\subsubsection{Intent Collusion Attack}
In case of \textit{intent} collusion attack, two malicious apps work together to leak the desired information. If a malicious app-A has the permission to read vehicle speed, location or other sensitive information but does not have any internet permission. Whereas malicious app-B has the internet permission but no other permission. In such case, Malicious app-A will send this sensitive \textit{intent} to malicious app-B where it will leak this information to a malicious server. As all of these actions can be performed in the background, identifying such apps is not an easy task.
\section{approach}
\vspace{-0.15cm}
In this paper, we present an approach  to identify information leakage in Android Auto/Android Automotive apps using taint flow analysis. The approach also provide the developer with mitigation tips in order to align with the Android coding standards \cite{mitigation}. As seen in Figure \ref{fig:txl}, The process starts by decompiling Android APK or DEX files into Java files using Jadx decompiler. Jadx (DEX to Java decompiler) is a command line interface/GUI tool to produce Java source code from Android APK and DEX files. The decompiled Java files then passed to a TXL interface which takes Java grammar and Java source code as input and finds data flow from sources to sinks.
TXL\cite{Cordy2011} is a source transformation language that is specifically designed to support computer software analysis and source transformation tasks. The TXL paradigm consists of parsing the input text into an Abstract Syntax Tree (AST), transforming the tree to create a new AST and unparsing the new tree to a new output text. Grammar and Transformation rules are specified in the TXL. All the TXL rules recursively apply to the parsed tree until it finds no matches. This way it transforms and analyzes all the statements and generates the output. 
The following steps will walk us through the approach process and provides examples on each step:
\begin{figure}[t!]
\centerline{\includegraphics[width=.49\textwidth]{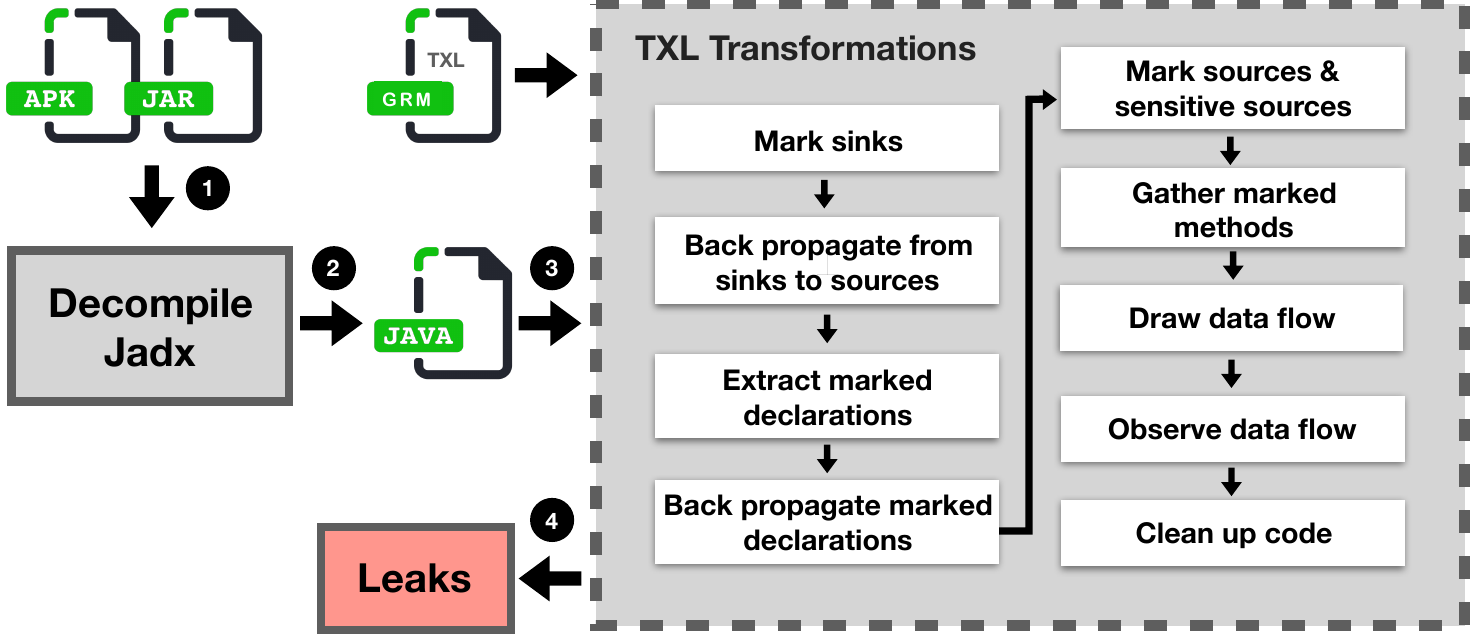}}
\caption{Approach}
\vspace{-0.5 cm}
\label{fig:txl}
\end{figure}
\subsubsection{Mark Sinks}
Listing \ref{lst2} demonstrates how the sinks are marked. This step globally defines classes and method names which are sources, sensitive sources and sinks. The goal is to use that information to  mark the sink statements in the code. These marked statements are the starting point for the back propagation step. This rule goes to each declaration or statement and match the sink method names. If a match is found it marks the statement with a sink tag and line number where it exists in the source code. In the case of \textit{sendBroadcast} it will only mark that as sink if it does not contain a custom permission parameter. The result of this rule can be seen in Listing \ref{lst3} (line 178-180). Once all the sinks are marked in the source code the process moves to the next rule.
\vspace{-0.2 cm}
\begin{lstlisting}[style=Android, caption={Define Sink TXL Rule},label={lst2}]
rule markSinks
    export sources [repeat stringlit]
        "Intent"
    export sensitiveSources [repeat stringlit]
        "Car" "CarInfoManager" "CarPropertyManager"
    export sinks [repeat stringlit]
        "sendBroadcast"
    replace $ [program]
        P [program]
    by
        P [markSendBroadcastSink]
end rule
\end{lstlisting}
\vspace{-0.2 cm}
\begin{lstlisting}[style=Android, firstnumber=175, caption={Mark Sink Output:},label={lst3}]
private void broadcastIntent () {
    Intent v1 = createIntent(key, value);
    ...
   <@\textcolor{red}{<sink 186>}@>
       ~sendBroadcast(v1);~
   <@\textcolor{red}{</sink>}@> }
\end{lstlisting}
\subsubsection{Back propagate from Sinks to Sources}
This rule extracts the variable names from sink methods and checks their usage on previous statements recursively. This process is called back propagates the flow from sink to source and finds the exact source which ends up in the sink. The marked statements can be seen in Listing \ref{lst4}, where process starts at (line 204), extracts variable name intent and this variable name matched with all the previous statements until variable declaration is found.
\subsubsection{Extract Marked Declarations}
This rule extracts all the marked declaration statements from the transformed code and exports them in a global variable. To search for any declarations which are constructed through a method call. As can be seen in Listing \ref{lst4} line 177, variable intent was created with a method call \textit{createIntent}.
\subsubsection{Back propagate marked declarations}
This rule utilizes the globally extracted declarations statements, matches the method name with those in the declaration statements. If a match is found this rule marks the return statement of that method, extracts the returned variable name and performs back propagation till it reaches the source. This rule ensures the declaration statement flow from method calls. As can be seen in Listing \ref{lst5}, method createIntent and all the relevant statements in this method are marked properly.
\vspace{0.2 cm}
\begin{lstlisting}[style=Android,firstnumber=175, caption={Back propagate from Sinks to Sources:},label={lst4}]
private void broadcastIntent () {
    ~<source 176>~
        Intent <@\textcolor{red}{intent}@> = createIntent (key, value);
    </source>
    <line 177>
        <@\textcolor{red}{intent}@>.setAction ("com.example.intent.broadcast");
    </line>
    *<sensitive_source 178>*
        CarInfoManager <@\textcolor{red}{carInfo}@> = (CarInfoManager) <@\textcolor{red}{car}@>.getCarManager(<@\textcolor{red}{Car}@>.INFO_SERVICE)};
    </sensitive_source>
    <line 179>
       <@\textcolor{red}{intent}@>.putExtra ("Battery Capacity", <@\textcolor{red}{carInfo}@>.getEvBatteryCapacity());
    </line>
    <line 180>
        <@\textcolor{red}{intent}@>.putExtra ("Connector Types", <@\textcolor{red}{carInfo}@>.getEvConnectorTypes());
    </line>
    <line 181>
        <@\textcolor{red}{intent}@>.putExtra ("Fuel Capacity", <@\textcolor{red}{carInfo}@>.getFuelCapacity());
    </line>
    <line 182>
        <@\textcolor{red}{intent}@>.putExtra ("Fuel Types", <@\textcolor{red}{carInfo}@>.getFuelTypes());
    </line>
    <line 183>
        <@\textcolor{red}{intent}@>.putExtra ("Manufacturer", <@\textcolor{red}{carInfo}@>.getManufacturer());
    </line>
    <line 184>
        <@\textcolor{red}{intent}@>.putExtra ("Model", <@\textcolor{red}{carInfo}@>.getModel());
    </line>
   <sink 186>
        sendBroadcast (<@\textcolor{red}{intent}@>);
   </sink>
   }
\end{lstlisting}
\vspace{-0.2 cm}
\begin{lstlisting}[style=Android,firstnumber=112, caption={Back propagate marked declarations:},label={lst5}]
<line 112>
    private Intent createIntent (String key, String value) {
        <source 115>
            Intent intent = new Intent ();
        </source>
        <line 116>
            intent.setAction ("com.example.automotiveintentexample.broadcast");
        </line>
        <line 117>
            intent.putExtra (key, value);
        </line>
        <line 129>
            return intent;
        </line>
    }
</line>
\end{lstlisting}
\subsubsection{Mark sources and sensitive sinks}
This rule checks for all the marked declaration statements in the code and replaces the line tag with \textit{source} tag if their data type is matched with the class name of source types. It will replace the line tag with  \textit{sensitive source} tag if matched with class name of sensitive source types. The output of this rule can be seen in Listing \ref{lst4} where line 176 \textit{Intent} is marked as \textit{source} and line 182 \textit{CarInfoManager} is marked as \textit{sensitive source} tag.
\subsubsection{Extract marked methods}
This rule checks for all the marked methods in the code and extracts them in a global variable so it can be later used for drawing data flow. As searching from marked method list increases performance.
\subsubsection{Draw Data Flow}
As the name implies this rule draws the data flow from sources to sinks. When all the relevant statements are marked, this rule searches for the source tag statement, extracts the variable name and recursively moves to the next line or jumps to method definitions to search for the next usage until it reaches the sink. 
For each relevant statement, it keeps track of the line number and adds that into the data flow variable so it can be passed to \textit{Observe} rule which prints out the output. When it no longer finds the usage of that variable in next statements it passes the flow to \textit{Observe Data Flow} rule which check each statement and produce warnings and leaks messages. Then it repeats this process for the next source tag statement.
\subsubsection{Observe Data Flow}
As mentioned in \textit{Draw Data Flow} rule, this rule gets a data flow variable which contains source code line numbers (e.g: 176 112 115...), then for each line number it extracts the statement and observes it, if that statement contains any leaks like \textit{sendBroadcast}, it reports the leak message and mitigation tip as seen in Listing \ref{lst6} (Line 8). If it finds that \textit{intent} object is getting any sensitive information it reports a warning message. As seen in Listing \ref{lst6} (Line 2-7).
\begin{lstlisting}[style=Android, basicstyle=\fontsize{8}{12}\selectfont, caption={Observe Data Flow:},label={lst6}]
Flow: 176 112 115 116 117 129 177 179 180 181 182 183 184 186
Warning: Source variable contains sensitive information - Line: 179
Warning: Source variable contains sensitive information - Line: 180
Warning: Source variable contains sensitive information - Line: 181
Warning: Source variable contains sensitive information - Line: 182
Warning: Source variable contains sensitive information - Line: 183
Warning: Source variable contains sensitive information - Line: 184
Leak: Send Broadcast leaking information to all the apps. Compliant solution requires usage of LocalBroadcastManager or sendBroadcast with custom permissions. - Line: 186
\end{lstlisting}
\subsubsection{Code Cleanup}
These rules remove all the unnecessary statements and method definitions from the output and removes all the markups to clean the output. So that when TXL prints the final transformation of source code, user can have the clean source code containing only relevant statements and methods as reported by the data flow.
\section{Related work}
\vspace{-0.15cm}
Inter Component Communication threats have attracted lots of research as they play an important role in detecting malicious apps that escalate their privileges by hijacking intents in Android apps. Tools like FlowDroid \cite{FlowDroid} perform taint analysis by statically analyzing .apk or .dex files, it depends on SuSi to collect sources and sinks through a machine learning approach which doesn't guarantee all the sources and sink coverage. Also, to perform FlowDroid analysis, we need to have a complete Android platform in which the app is developed, making it harder to perform analysis. As we learned from the experiment we did in the evaluation section, FlowDroid doesn't recognize Android Auto, or Android Automotive libraries as such it's general analysis is not useful for the work we do in this paper.

A recent work, SIAT tool \cite{hu2020siat}, provides dynamic analysis to test apps for information leakage. The approach  modifies android framework by adding a monitor and analyzer to detect leaks. To use SIAT, users need to configure their devices with SIAT custom android framework, then users need to write an automation script for each app. SIAT is only able to detect leaks when information leaves an app and is received by another. In case no app receives such information it will not report that as a potential leak. So it has chances of missing leaks if it does not find a suitable app pair. Whereas, in our case, we analyze  every single app and report for potential information leakage, as this will provide a better coverage for potential information leakage even if the app was not tested when interacting with other apps.

Another static analysis approach, SDLI \cite{8456010}, built on top of a commercial static analyzer Julia, it generates XML reports for each inward and outward intent and compares XMLs reports for intent leakage. SDLI reports for leaks whenever any sensitive information leaves the app and reaches another app. They are not considering secure communication or mentioned anything related to app permissions. In our case, we report information leakage and provide mitigation tips to the developers, so they can ensure the ICC communication happen in a secure manner and \textit{intent} from an app is received by an intended app or broadcasted locally in the same app. Hence, keeping the app to app communication secure.


Recent work by Amit et al \cite{doi:10.1002/spe.2698} also explores the unprotected intent, In their work they are using null fuzzing of intent to generate blank intents with no data to see which activities and services of apps became active due to it. Their approach helps in finding vulnerable communication component. As they report, due to intent fuzzing, many of the infotainment services became unresponsive and many apps crashed due to missing information. Their approach helps us understand the importance of secure app to app communication specially in the automotive domain. Unresponsive infotainment systems can lead to driver distraction which impacts the safety, security of the driver as well as vehicles around it.

Different from all the above techniques, in addition to detecting and reporting information leakage, we provide mitigation tips.
\section{Experiment and Evaluation}
\vspace{-0.15cm}
This section presents the experiment and evaluation of our tool based on the following datasets. 
We downloaded 236 Android Auto apps manually from Google play store to perform static analysis and we also downloaded 40 reversed engineered apps provided and used in \cite{doi:10.1002/spe.2698}, all these reversed engineered apps were in jar format. 

For our experiment, we decompiled all the apk and jar files into Java code using Jadx decompiler and then performed our analysis. As seen in Table \ref{table:results}, our tool performed analysis on 273 apps in which it was able to find 402 sendBroadcast leaks from 121 apps. We inspected the flow manually and also verified with an automated script that the flow contains \textit{intent} as source and \textit{sendBroadcast} as sink and it confirmed our results that all those cases were true positives.
\begin{table}
\centering
\caption{Analysis results of our approach and FlowDroid.}
\vspace{-0.2cm}
\begin{tabular}{|l|l|l|l|}
\hline
\textbf{Tool} & \textbf{Total Apps} & \textbf{Apps with leaks} & \textbf{Leaks} \\ \hline
AAVD          & 273                 & 121                      & 402            \\ \hline
FlowDroid     & 204                 & 33                       & 74             \\ \hline
\end{tabular}
\label{table:results}
\end{table}
To compare our tool with state of the art approaches, we tried to compare with SIAT \cite{hu2020siat}, as they have made their code public in Github, but hardly any information about its usage. Also, they are using dynamic analysis, for us to compare we need to implement some automation script for each app. For SDLI \cite{8456010} and the approach presented by Amit et al \cite{doi:10.1002/spe.2698}, we were not able to find their tool publicly available, this maybe due to the fact that both tools were built on top of Commercial Julia Analyzer, so they were not able to distribute it publicly.


As such, we were only able to compare with the open source tool, FlowDroid\cite{FlowDroid}. As can be seen in Table \ref{table:results}, FlowDroid was able to perform analysis on 204 apps out of 236 Android Auto apps downloaded manually and it completely failed to work on the 40 reversed engineered apps which were in jar format. From 204 apps, it found 74 \textit{sendBroadcast} leaks in 33 apps. We wrote a script to find overlapping leaks between FlowDroid and our approach and found no overlapping leaks. 

The difference came from how these two approaches work. As for FlowDroid, we need to provide it with a file which defines what should be considered as sources and sinks. These sources and sink definition is very granular in FlowDroid, it needs to know exactly which method is source and which method is sink for each class, and with rapidly changing Android SDK these sources and sinks file is not up to date. However, our approach is general, we are considering \textit{intent} as a source, \textit{CarInfoManager}, \textit{CarPropertyManager} as sensitive source and \textit{sendBroadcast} as sink. We are considering all \textit{sendBroadcast} as leaks unless it’s being called with custom permission which makes it secure. FlowDroid, is not designed for infotainment apps, hence it does not provide checks specific to such apps.
\begin{lstlisting}[style=Android, caption={Data Flows Extracted by our tool},label={lst7}]
App: ~<com_jacapps_wilfm_apk>~
File: `TweetUploadService.java`
Statements: {
    126: Intent intent3 = new Intent(UPLOAD_FAILURE);,
    127: intent3.putExtra(EXTRA_RETRY_INTENT, intent2);,
    128: intent3.setPackage(getApplicationContext().getPackageName());,
    129: sendBroadcast(intent3);
}
App: ~<com_abc_abcnews_apk>~
File: `GoogleNowAuthorizationService.java`
Statements: {
    22: Intent intent2 = new Intent(Constants.BROADCAST_ACTION);,
    26: intent2.putExtras(extras);,
    33: intent2.putExtra("authCode", NowAuthService.getAuthCode(this, Constants.SERVER_CLIENT_ID));,
    43: intent2.putExtra(Constants.ACCESS_TOKEN_EXTRA, e5.getAccessToken());,
    45: intent2.setPackage(getPackageName());,
    46: sendBroadcast(intent2);,
    49: intent2.setPackage(getPackageName());,
    50: sendBroadcast(intent2);}
\end{lstlisting}
Listing \ref{lst7} present some of the cases found by our tool. As seen in Line 4, Intent \textit{UPLOAD\_FAILURE} is found in 31 apps, if this intent is broadcasted then it will surely be received by all the 31 apps. However, from statements at (Line 4-7) it seems like this information is only required within the same app context. Using \textit{LocalBroadcastManager} in this place is the required solution. 

App \textit{com\_abc\_abcnews} at (line 9), is broadcasting an authorization token to all the apps. (Line 12-19), from filename at (line 10) we know this authorization token belongs to \textit{GoogleNowAuthorizationService}. Any malicious app with this information can do serious damages to the user. 
\section{Conclusion and Future work}
\vspace{-0.15cm}
In the paper, we presented ICC threat models and their severity with respect to Android Automotive and how it can be used to distract drivers, steal their private information. We also present the design and implementation of our approach to detect ICC leaks and generation of warning messages to mitigate it. Our approach is still in its early phase as we progress to improve it further by considering multiple source files analysis so we could discover leaks between multiple source files. We also want to include more ICC sinks that can leak \textit{intent} and to provide a secure solution to prevent it. 

In our current implementation, any instance of \textit{sendBroadcast} which contains a permission parameter is not flagged as a leak, so we want to enhance the approach further by checking Android permissions, since Android permissions are requested by the majority of apps, and in that scenario such broadcast is still not secure. Having custom permission ensures that only dedicated apps can communicate securely. 

We also want to enhance our tool to check for broadcast listeners and to prevent it from listening to insecure broadcasts. Secure implementation needs to happen from both Intent generation as well as intent listening endpoints. Finally, we want to implement an app comparison feature to detect ICC leaks among app pairs.
\balance
\bibliographystyle{IEEEtran}
 \bibliography{Abdul_SCAM.bib}
\end{document}